# Optical frequency transfer via 146 km fiber link with $10^{-19}$ relative accuracy

G. Grosche,[1,*] O. Terra,[1] K. Predehl,[1,2] R. Holzwarth,[2] B. Lipphardt,[1] F. Vogt,[1] U. Sterr,[1] and H. Schnatz[1]

[1]*Physikalisch-Technische Bundesanstalt, Bundesallee 100,*

*D-38116 Braunschweig, Germany*

[2]*Max-Planck-Insitut für Quantenoptik, Hans-Kopfermann-Str. 1*

*D-85748 Garching, Germany*

[*]*Corresponding author: gesine.grosche@ptb.de*

We demonstrate the long-distance transmission of an ultra-stable optical frequency derived directly from a state-of-the-art optical frequency standard. Using an active stabilization system we deliver the frequency via a 146 km long underground fiber link with a fractional instability of $3\times10^{-15}$ at 1 s, which is close to the theoretical limit for our transfer experiment. The relative uncertainty for the transfer is below $1\times10^{-19}$ after 30 000 seconds. Tests with a very short fiber show that noise in our stabilization system contributes fluctuations which are two orders of magnitude lower, namely $3\times10^{-17}$ at 1 s, reaching $10^{-20}$ after 4000 s.

Ultra-precise optical frequency standards (OFSs) have reached a relative instability and accuracy below $10^{-16}$ [1]. This puts increased demands on methods for disseminating frequencies and for comparing standards, which are usually located in different laboratories. Long-distance transmission of either an amplitude modulated carrier [2] or an ultra-stable optical carrier frequency [3-8] over tens or hundreds of kilometers using standard telecommunication fiber is a powerful, relatively new alternative to satellite-based methods [9]. Hong *et al.* [10] recently demonstrated the first remote frequency measurement of the $^{87}$Sr optical clock transition via a 120 km fiber link; its uncertainty was limited by the relative instability of the transmitted frequency (derived from an H-maser) of $4.7\times10^{-13}$ $(\tau/s)^{-1/2}$, where $\tau$ is the measurement time. As part of a German research fiber network, we have set up a fiber link bridging almost 70 km geographical distance between PTB Braunschweig and the Leibniz Universität Hannover (LUH), to directly connect ultra-stable OFSs [11,12,13], with a relative instability near $10^{-15}$ at 1 s, in the two institutions. The link consists of two parallel fibers. To characterize the link, we connected them, forming a 146 km loop with the input and remote ends located in the same lab at PTB.

A key element for frequency transfer via optical fiber is an interferometric fiber-stabilization system to suppress phase fluctuations arising along the fiber link. In the past, intrinsic noise in this system has been identified as limiting the overall performance for long measurements. Williams *et al.* [6] reported this instability floor at $2\times10^{-17}$ (1 s), scaling with $1/\tau^{1/2}$, and reaching $2\times10^{-19}$ at 10 000 s. We address this limitation and report two results. First, we designed a low noise stabilization system, which delivers an optical frequency through a short fiber with a relative instability and uncertainty below $3\times10^{-17}$ at 1 s, scaling with $\sim1/\tau$, and reaching $10^{-20}$ at 4000 s. Secondly, we stabilized the 146 km fiber link which connects PTB and LUH: we delivered an ultra-stable frequency to the remote end with a relative instability of $3\times10^{-15}$ at 1s, scaling as $1/\tau$ up to 30 000 s; the relative accuracy is better than $10^{-19}$. The results are a factor hundred better than reported in [10]; compared to previous reports for a comparable fiber link [7,8] and for the interferometer [4,6,7,8], we have improved the long-term instability by more than a factor of five and ten, respectively.

As light source, we phase-stabilize a fiber laser at 1542 nm (Koheras Adjustik) to a cavity stabilized optical clock laser at 657 nm with a line-width of about 1 Hz, using a fs-frequency comb as transfer oscillator [14]. The clock laser [12] is part of the Ca neutral atom frequency standard at PTB; its frequency stability is mapped to 1542 nm. Thus the stabilized fiber laser has a fractional frequency instability of $2\times10^{-15}$ (1s) and a coherence length > 10 000 km. We inject ~3 mW optical power into the fiber link, keeping below the threshold for stimulated Brillouin scattering.

The fiber link and compensation interferometer are shown in Fig. 1. Two fibers, F1 and F2 (in the same cable) connect our lab at PTB with the computer centre (RZ-H) of the LUH. F1 and F2 were connected at RZ-H, resulting in a fiber link starting and ending at PTB. The single-pass link length is 146 km and incorporates one home-built bi-directional low noise erbium-doped fiber amplifier to partially compensate the 43 dB single-pass optical attenuation of the fiber.

Mechanical perturbations and temperature changes lead to optical path length fluctuations, which appear as a time-varying Doppler shift of the transmitted light frequency. These fluctuations are compensated with a commonly used scheme [3-8] based on Ma *et al.* [15]: ideally, the entire fiber link is treated as one arm of a Michelson interferometer, which is phase-stabilized to a reference arm. As shown in Fig. 1, we detect a double pass signal ("in-loop", at PD1) and derive a correction signal which controls the frequency and phase of an acousto-optic modulator (AOM1) at the fiber input. AOM2 serves to distinguish light reflected by the Faraday mirror at the remote end from spurious reflections along the fiber. The stabilized link is characterized by measuring a beat signal between out-going laser light and light at the remote end (remote or "out-of-loop" signal, at PD2). Any residual noise in the reference arm, and any uncompensated optical paths contributing to the remote signal, give rise to an instability floor which is *independent* of the fiber link itself.

The signal at the remote end (PD2) was tracked with a bandwidth $f_b > 10$ kHz using a phase locked loop, and counted with a zero dead-time totalizing counter which acts as a "Π-estimator" [16]. From this we calculate the Allan deviation (ADEV) as our measure of relative frequency instability. As discussed in [7,16], other measures of instability [4-8,10], such as the *modified* Allan deviation (ModADEV), may give much smaller numerical results for signals dominated by white phase noise; such signals are common for stabilized fiber links.

In our lab, unstabilized optical fiber with lengths of 1 m … 20 m showed phase noise $S_\phi(f)$ of $10^{-3}$ … $10^{-1}$ rad$^2$/Hz $\times f^{-2}$, and an ADEV of up to $2\times10^{-15}$ (1 s). The ADEV for an unstablized 20 m patchcord is shown in Fig. 2 (open green circles). We tested different designs for the interferometer, using off-the-shelf fiber-optic components and minimizing uncompensated fiber in a compact set-up. A key idea is to incorporate components in such a way that the in-loop signal experiences *exactly* twice the perturbations that act on the remote signal; as a simple example, AOM1 and AOM2 have effectively become part of the stabilized fiber link. The reference arm for each heterodyne beat is kept short. The interferometer was covered with a lid to reduce air currents causing temperature drifts.

To characterize the compensation interferometer, we measure a fiber link of zero length, without EDFA. We connect the output of AOM1 with the input of AOM2, and denote as *interferometer noise floor* the measured phase noise and ADEV of the signal obtained at the remote end, when the in-loop signal is used for stabilization. The phase noise is approximately $S_\phi(f) = 10^{-7}$ rad$^2$/Hz $\times f^{-1}$, for Fourier frequencies between 1 Hz and 1 kHz (red dashed line in Fig. 3 shows $S_\phi(f) \times 1000$). The ADEV was $3\times10^{-17}$ at 1 s, reaching $10^{-20}$ after 1 hour (red open triangles in Fig. 2); beyond 10 000 seconds, we suspect a flicker floor of several $10^{-21}$. Averaging over the whole dataset of >220 000 s, the mean remote frequency differed from its nominal value by 0.68 µHz, or $3.5\times10^{-21}$. We compare this with the best previously reported data [6]: an optimized free-space interferometer showed a ModADEV of ~ $2\times10^{-17}$ (1 s), scaling with $1/\tau^{1/2}$ and reaching a level near $10^{-19}$ after 10 000 seconds [6]; the $1/\tau^{1/2}$-behaviour means that ADEV and ModADEV are related by ADEV = ModADEV $\times 2^{1/2}$ [16]; Fig. 2 (dotted red line) shows ADEV data calculated from [6] in this way.

The result for the stabilized 146 km link is shown as full black squares in Fig. 2. The ADEV is approximately $3\times10^{-15}/(\tau/s)$, reaching $10^{-18}$ after one hour and $1\times10^{-19}$ after 32 000 seconds. Π-counter data recorded with a gate time of 20 ms gives a ModADEV of $5\times10^{-16}$ at 1 s. Figure 3 shows the phase-noise spectra for the link when stabilized ($S_\phi^{stab}(f)$) and unstabilized ($S_{\phi,}^{free}(f)$); the noise suppression is close to the theoretical limit [6], of $S_\phi^{stab}(f) / S_\phi^{free}(f) \geq 4\pi^2/3 \times (f\tau_d)^2$ (within the servo bandwidth); the single pass delay time $\tau_d$ is 0.7 ms in our case. At $f = 1$ Hz the equation gives a noise suppression of 52 dB; we observe ~50 dB. Stabilized, the square root of the *integrated* phase noise up to 10 kHz is 2.2 radian, corresponding to an ADEV of $3\times10^{-15}$ at 1 s, in agreement with counter measurements. If we integrate the phase noise only up to 10 Hz [7], we obtain 0.35 radian, equivalent to a 1 s instability of $4\times10^{-16}$.

We also compared the transmitted optical frequency to an *independent* optical clock laser [13] at the remote end: we thus compared two ultra-stable optical clock lasers via a 146 km fiber link for the first time. The ADEV was $5\times10^{-15}$ at 1 s; for further experimental details see [17].

The accuracy of the frequency transfer depends on systematic shifts: shifts due to the interferometer set-up and signal detection were below $10^{-20}$ (see above), but polarization mode dispersion and external magnetic fields may cause time-dependent non-reciprocity along the link. We estimate such shifts to be smaller than the statistical uncertainty. This is confirmed by three long measurements lasting $T_{max}$= 54 000 s, 50 000 s and 164 000 s, respectively: the mean remote frequency for the 146 km link, averaged for each data-set, differed from its nominal value by 7 µHz (fractional deviation $4\times10^{-20}$), 1.8 µHz ($1\times10^{-20}$) and -1.2 µHz ($-6\times10^{-21}$), respectively. The fractional deviation is therefore one order of magnitude smaller than the statistical uncertainty, for which we take the last ADEV value (at $\tau \approx T_{max}/4$) for each data set, at $3\times10^{-19}$, $2\times10^{-19}$ and $1\times10^{-19}$, respectively. Overall, the frequency transfer via fiber is found to be much more stable and accurate than currently available clocks.

We have demonstrated a high fidelity long-distance transmission of an ultra-stable optical frequency derived directly from a state-of-the-art optical frequency standard. This allows remote laser characterization and remote precision frequency measurements. We aim to extend the fiber link to a 1000 km network connecting LUH, PTB, Universität Erlangen and the Max-Planck-Institute for Quantum Optics near Munich: with comparable fiber properties, we expect an ADEV of $\sim 5\times10^{-14}/(\tau/s)$ if a 900 km-link from Braunschweig to Munich were stabilized in a single stretch. Our improved interferometer noise floor allows very low instability frequency distribution over short distances: we calculate an ADEV below $5\times10^{-17}/(\tau/s)$, reaching $\sim10^{-20}$ after one hour, for distances up to ~ 6 km. Given the considerable effort in developing ultra-stable laser sources [12,18,19], recent demonstrations of cavity stabilized lasers with a 1 s-instability below $6\times10^{-16}$ [20], and the proposal for a mHz-linewidth laser [21], this is attractive for an "on-campus" frequency distribution.

The transfer of a known, ultra-stable frequency with a relative instability in the $10^{-14}$ to $10^{-19}$ range over distances up to hundreds of kilometers between any "local" and "remote" laboratory connected by optical fiber makes precision frequency metrology available to a new class of end users. With commercially available fs-frequency combs, the delivered optical frequency may be mapped to any optical [1,14] or microwave [22-24] frequency in the remote lab. Thus fiber-based frequency transmission can be used for the remote frequency measurement of optical [10] or rf-transitions, the remote generation of precision microwave signals, remote characterization of laser sources and remote spectroscopy and sensing.

This work was supported by the Deutsche Forschungsgemeinschaft (grant no. SFB 407), and the Centre for Quantum Engineering and Space-Time Research (QUEST). We thank Thomas Udem and Christian Lisdat for helpful comments.

Figure captions

Fig. 1. (Color online): Schematic set-up of active fiber noise compensation for a 146 km fiber link. Laser: fiber laser stabilized to optical clock laser via fs-comb [14], OC: optical circulator, AOM: acousto-optic modulator, VCO: voltage-controlled oscillator, PD1: double-pass (in-loop) photo-detector, PD2: remote (out-of-loop) photo-detector, EDFA: erbium-doped fiber amplifier, FM: Faraday mirror.

Fig. 2. (Color online): Relative frequency instability of the *remot*e signal given by the Allan deviation (ADEV) using Π-type counters with a recording bandwidth > 10 kHz. Open red triangles: interferometer noise floor (obtained by shorting the fiber link, *stabilized*); full black squares: 146 km fiber link, *stabilized*; open green circles: 20 m of *unstabilized* fiber. Previously reported data for comparison: $Al^+/Hg^+$ clock comparison (solid magenta line: [1]); GPS based frequency transfer (dashed blue line: [9]); reported noise floor of an optimized free-space interferometer (dotted red line: ModADEV in [6] converted to ADEV [16])

Fig. 3. (Color online): (a) Phase noise spectral density $S_\phi(f)$ ($rad^2/Hz$) for beat signals detected at the remote end. Dashed thin red line: $1000 \times S_{\phi,}^{1m\text{-stab}}$ (interferometer noise floor measured by stabilizing 1m patchcord); thin green line: $S_{\phi,}^{free}$ (unstabilized 146 km link); thick black line: $S_{\phi,}^{stab}$ (146 km link with fiber stabilization). The theoretical limit due to "delay-unsuppressed noise" [6] was found to coincide with stabilized 146 km data. (b) Square root of phase noise integrated up to frequency *f*, and timing jitter for the 146 km link. Thin green line: unstabilized; thick black line: stabilized. The servo-bump is visible near 300 Hz. Phase noise around *f* = 15 Hz contributes roughly two thirds of the total phase noise within the servo-bandwidth.

Figure 1

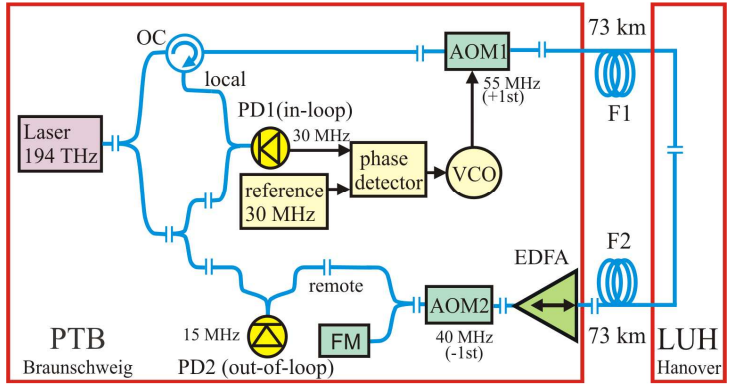

Figure 2

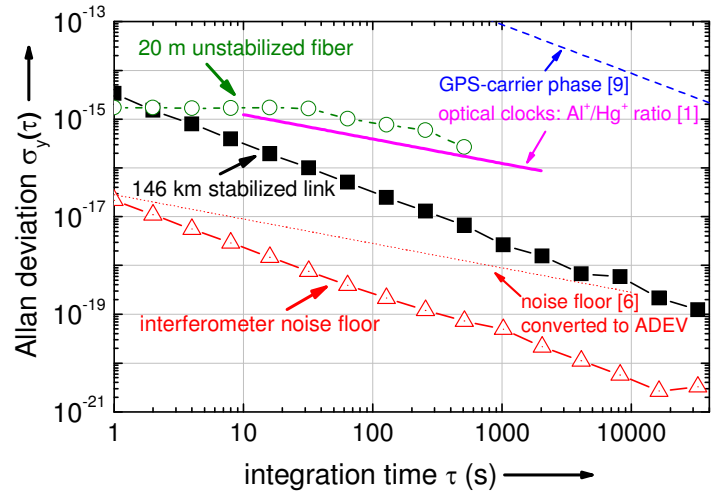

Figure 3

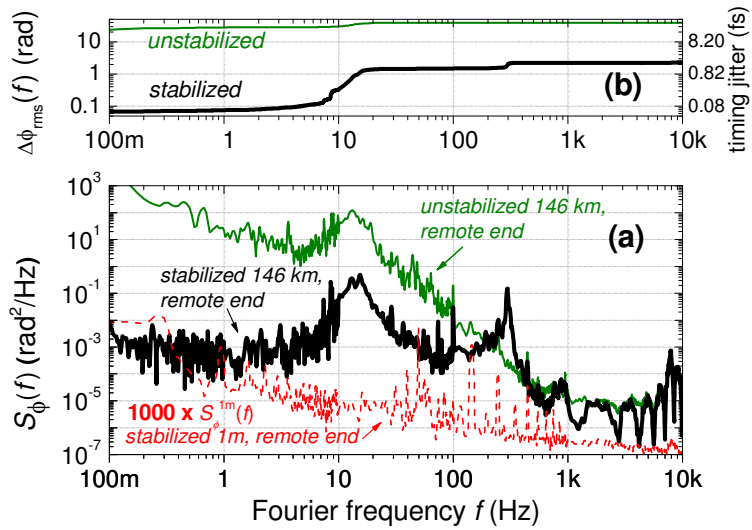